\documentclass[reprint,amsmath,amssymb,aps,superscriptaddress,]{revtex4-1}

\usepackage{graphicx}
\usepackage{subfigure} 
\usepackage{dcolumn}
\usepackage{bm}
\usepackage{color,soul} 
\usepackage{upgreek}
\usepackage[percent]{overpic}

\begin{document}

\preprint{APS/123-QED}

%
%
\title{Topological origin of strain induced damage of multi-network elastomers by bond breaking}

\author{Yikai Yin}
\affiliation{
 Department of Materials Science and Engineering, Stanford University, Stanford, CA 94305
}
\author{Nicolas Bertin}
\affiliation{
 Department of Mechanical Engineering, Stanford University, Stanford, CA 94305
}
\author{Yanming Wang}
\affiliation{
 Department of Materials Science and Engineering, Stanford University, Stanford, CA 94305
}
\affiliation{
 Department of Mechanical Engineering, Stanford University, Stanford, CA 94305
}
\author{Zhenan Bao}
\affiliation{
 Department of Chemical Engineering, Stanford University, Stanford, CA 94305
}
\author{Wei Cai}
\affiliation{
 Department of Materials Science and Engineering, Stanford University, Stanford, CA 94305
}
\affiliation{
 Department of Mechanical Engineering, Stanford University, Stanford, CA 94305
}

\date{\today}

\begin{abstract}
Elastomers that can sustain large reversible strain are essential components for stretchable electronics.
%
The stretchability and mechanical robustness of unfilled elastomers can be enhanced by introducing easier-to-break cross-links, e.g. through the multi-network structure, which also causes stress-strain hysteresis indicating strain-induced damage.
However, it remains unclear whether cross-link breakage follows a predictable pattern that can be used to understand the damage evolution with strain.
Using coarse-grained molecular dynamics and topology analyses of the polymer network, we find that bond-breaking events are controlled by the evolution of the global shortest path length between well-separated cross-links, which is both anisotropic and hysteretic with strain.
These findings establish an explicit connection between the molecular structure and the macroscopic mechanical behavior of elastomers, thereby providing guidelines for designing mechanically robust soft materials.
\end{abstract}

\pacs{Valid PACS appear here}
\maketitle





Elastomers are fundamental building blocks for stretchable electronics, which enables novel wearable and biological applications \cite{chortos2014skin, wagner2012materials}.
However, maintaining mechanical robustness under large cyclic strain is still a major challenge.  
While a common strategy to improve the mechanical properties of the elastomers is to add filler particles, these particles may interfere with the electronic properties~\cite{flandin2001interrelationships}, and impose compatibility constraints for processing~\cite{heinrich2002reinforcement,mark2007rubberlike}.
An alternative approach is to add cross-links (covalent or non-covalent) with tailored properties to polymer networks~\cite{gong2014materials}, enabling the mechanical properties to be tuned independently of the electronic properties~\cite{oh2016intrinsically}, and taking advantage of the versatility of modern chemical methods~\cite{appel2012supramolecular,kang2018tough}.

Recent experiments have shown that elastomers with a multi-network structure can exhibit high stretchability and mechanical robustness simultaneously~\cite{ducrot2014toughening,millereau2018mechanics}.
Based on an approach similar to that in tough hydrogels~\cite{sun2012highly,webber2007large}, a double network (DN) is obtained by infiltrating a cross-linked single network (SN) with monomers followed by polymerization. A triple network (TN) is then obtained by repeating the process on the DN~\cite{ducrot2014toughening}.
Consequently,  these multi-network elastomers exhibit a strong stress-strain hysteresis under cycling loading, similar to the Mullins effect for filled SN elastomers~\cite{mullins1969softening,diani2009review}, except that there are no filler particles in these materials.
The strong hysteresis and improved mechanical properties are attributed to the fact that the polymer chains on the first network 
are pre-stretched by the insertion of subsequent networks before loading is applied~\cite{matsuda2016yielding}.
As a result, the cross-links on the first network become easier to break upon loading, which has been confirmed \textit{in situ} using chemoluminescent cross-linkers that emit light upon breaking~\cite{ducrot2014toughening}. 
However, the mechanism controlling how many and which cross-links should break at a given strain is not yet understood, and a microstructural parameter for the polymer network that explains the hysteretic stress-strain behavior is still missing.
More generally, existing constitutive models of elastomers~\cite{diani2009review,bacca2017model} are either phenomenological or based on assumptions not tested against the more fundamental molecular simulation models.
The identification of a microstructural damage parameter for bond breaking is therefore of critical importance for developing physics-based models that can guide the design of novel supramolecular elastomers with tunable mechanical properties \cite{cordier2008self,burnworth2011optically}.

Here we use coarse-grained molecular dynamics (CGMD) simulations to establish the connection between bond breaking and stress-strain hysteresis through the microstructural evolution of the polymer network.
While CGMD simulations have been previously applied to elastomers~\cite{davidson2016nonaffine,li2016molecular}, there have been few studies on bond-breaking events under large deformation, or on multi-network elastomers.
Our CGMD simulations predict stress-strain hysteresis and strain-induced bond-breaking events very similar to the experimental observations~\cite{ducrot2014toughening}. Furthermore, the CGMD simulations reveal that the strain-induced damage is anisotropic, which indicates that bond-breaking events occur in a non-random fashion.
We analyze the topology of the molecular chain networks by computing the \emph{shortest paths} (SPs) between well-separated cross-links and find that the length distribution of SPs is both anisotropic and hysteretic with strain. 
We demonstrate that the average SP length is the key microstructural feature that connects molecular level bond-breaking events to the macroscopic mechanical response, and that can serve as a foundation for physics-based constitutive models of damage evolution in highly stretchable elastomers.

\begin{figure*}[tp]
\centering
\includegraphics[width=\linewidth] {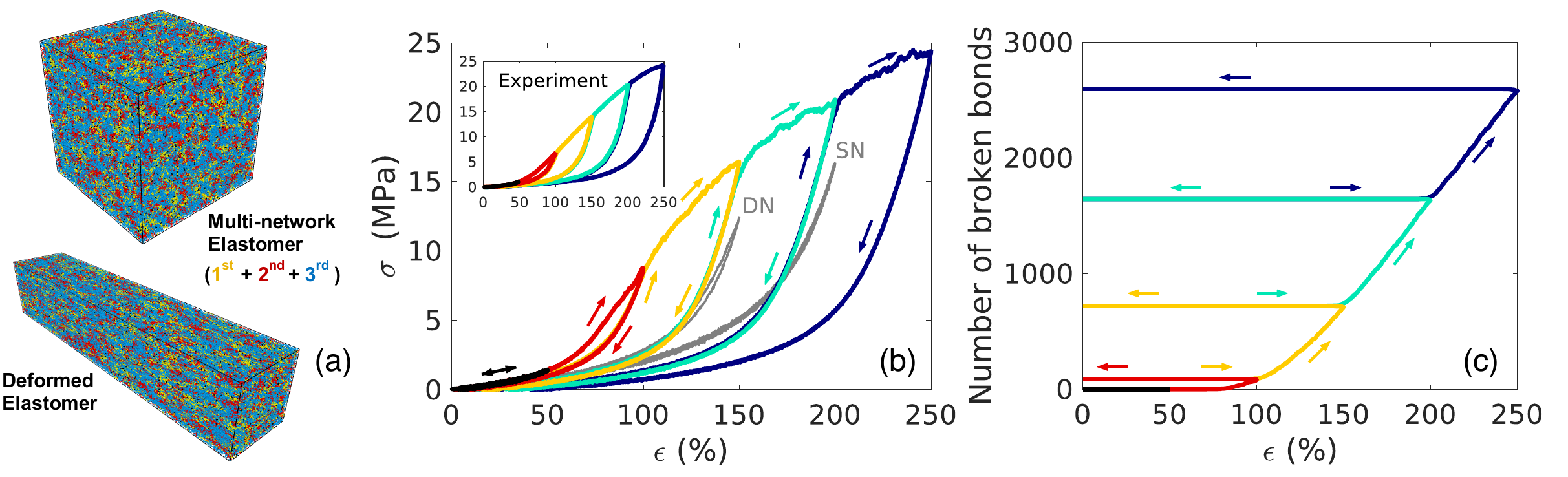}
\caption{(a) Snapshots of the TN elastomer configuration before and after deformation. (b) Stress-strain ($\sigma$-$\epsilon$) curves during loading-unloading cycles (indicated by arrows) with increasing maximum strain. (Inset) Similar $\sigma$-$\epsilon$ responses observed in experiments \cite{ducrot2014toughening}. (c) Evolution of the number of broken bonds during loading and unloading.
} 
\label{fig:fig1}
\end{figure*}


We perform CGMD simulations using LAMMPS~\cite{plimpton1995fast} on a bead-spring model~\cite{kremer1990dynamics} subjected to periodic boundary conditions.
%
%
The backbone interactions between neighboring beads on the same chain are modeled by the finite extensible nonlinear elastic (FENE) bonds. Non-bonded interactions between beads are modeled by a Lennard-Jones potential.
After equilibration with a two-step protocol~\cite{sliozberg2012fast}, 
cross-links between nearby beads on different chains are randomly added, turning the polymer melt into an SN elastomer. The cross-links are modeled using a quartic potential~\cite{stevens2001interfacial,ge2013molecular}, and are not allowed to re-form once broken. 
The simulation protocol for creating the multi-network structures is analogous to the experimental procedures of swelling, polymerization, and cross-linking~\cite{ducrot2014toughening}.
DN and TN structures are obtained by inserting more polymer chains to the SN, and adding cross-links between these new chains after equilibration. The presence of the lightly cross-linked subsequent networks isotropically stretches the chains in the first network as the volume of the elastomer expands. 
The SN, DN and TN models are then subjected to cyclic tensile strains in the $x$ direction with conservation of volume. See Supplemental Material (SM) for model details.
%

Fig.~\ref{fig:fig1}(a) shows simulation snapshots of the TN elastomer before and after deformation.
The predicted stress-strain curves are shown in Fig.~\ref{fig:fig1}(b).
Five consecutive loading-unloading cycles were performed, with the maximum strain increasing from $50\%$ to $250\%$. 
%
The stress-strain curves of the TN exhibit strong hysteresis and show remarkable agreement with experimental measurements reported in~\cite{ducrot2014toughening} (inset).
The stress during unloading follows a lower branch than that during loading. During reloading, the stress initially follows the lower branch and the stress-strain curve becomes reversible as long as the previous maximum strain is not reached. After stretching beyond the previous maximum strain, the stress follows the upper branch again during loading, and follows a new lower branch during unloading, all in agreement with the experimental measurements~\cite{ducrot2014toughening}. 
On the other hand, the stress-strain curves for the SN and DN (grey lines) exhibit minimal hysteresis at this level of strain, although they do show significant hysteresis at higher strain (see SM).

Fig.~\ref{fig:fig1}(c) shows the number of broken bonds as a function of strain in the TN, also consistent with experimental observations~\cite{ducrot2014toughening}. Specifically, bond breaking occurs during the initial loading beyond a critical strain (about $50\%$ for TN), and no bond breaking occurs during unloading and reloading until the previous maximum strain is exceeded. 
%
%
Furthermore, no bond-breaking events occur in the lightly cross-linked second and third networks, also consistent with experiments.

The success of our CGMD simulations in capturing the stress-strain hysteresis (Mullins effect) of unfilled elastomers with breakable cross-links provides an opportunity to answer the following fundamental question: which feature of the elastomer network governs bond-breaking events and the mechanical response? In other words, is there a {\it quantifiable microstructural parameter} that controls the strain-induced damage evolution?

\begin{figure}[tp]
\centering
\includegraphics[width=\linewidth] {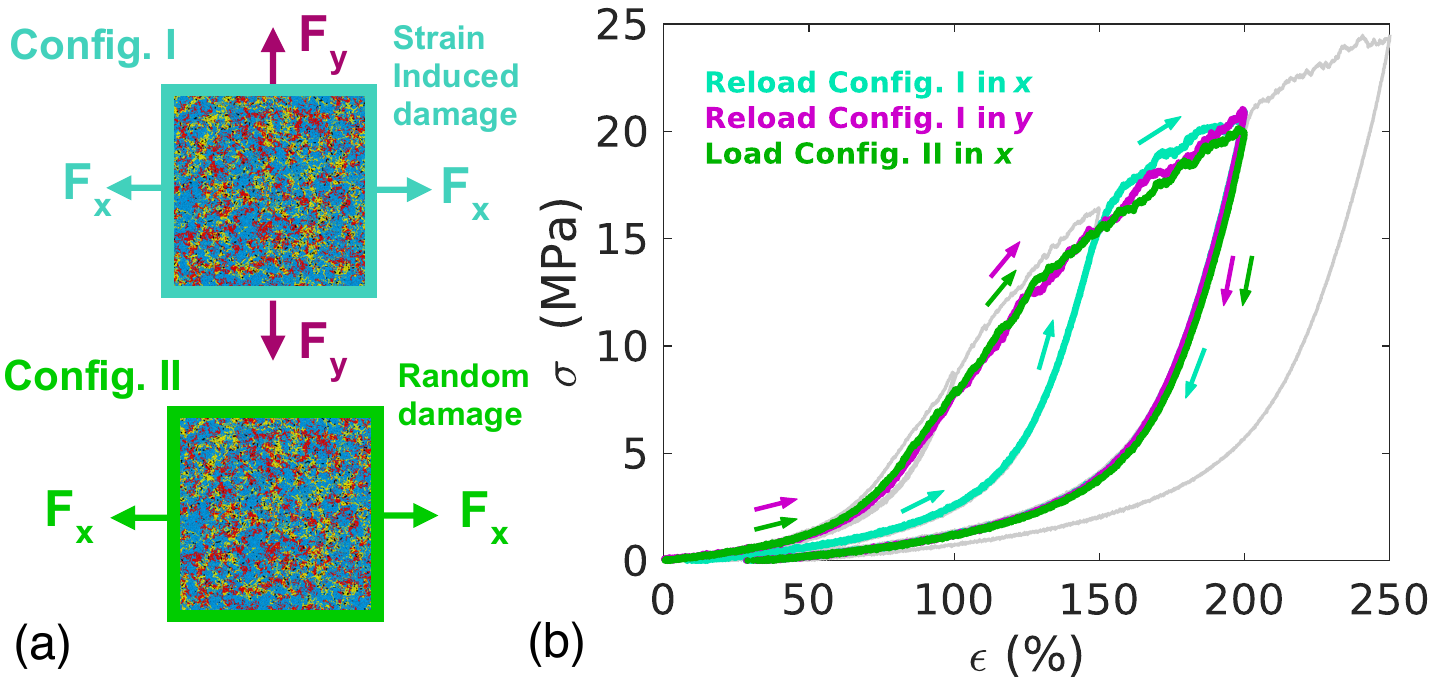}
\caption{ (a) Loading paths of different configurations with the same number of previously broken cross-links (see text). (b) $\sigma$-$\epsilon$ curves under different loading paths depicted in (a).} 
\label{fig:fig2}
\end{figure}

A natural candidate for the controlling microstructural parameter is the number (or density) of the cross-links, as suggested by the clear correlation between Figs.~\ref{fig:fig1}(b) and (c). However, we find that the number of cross-links alone does not govern the mechanical response, as two configurations with the same amount of cross-links but different loading histories can have very different behaviors. 
As shown in Fig. 2(a), the first configuration (I) is obtained from the original sample by stretching to $150\%$ strain in $x$, followed by unloading; 
and the second configuration (II) is obtained by randomly removing cross-links from the original sample.
%
Fig.~\ref{fig:fig2}(b) shows the stress-strain curves for the two configurations subjected to a $200\%$ strain cycle in $x$. While config. I with strain induced damage initially follows the lower unloading branch of the previous cycle (as expected), config. II with random damage follows the upper branch from the beginning. 
Since both configurations have the same number of cross-links, it follows that the cross-links do not break randomly during the initial stretching (to $150\%$ strain).
%
%
Furthermore, we find that after the initial stretching along $x$ the elastomer (config. I) becomes anisotropic .
Fig.~\ref{fig:fig2}(b) shows that if config. I is stretched along the $y$ direction, the stress-strain response is very similar to that of config. II (in which cross-links were removed randomly).
%
%
This result reveals that the strain-induced damage is anisotropic, and cannot be described by a scalar, such as the number of broken cross-links.

In light of these findings, we propose that a suitable microstructural parameter for strain-induced damage in unfilled elastomers needs to satisfy three criteria: (i) be {\it hysteretic} with strain, (ii) be {\it anisotropic}, and (iii) {\it controls the stress-strain} response.
The number of cross-links satisfies the first criterion but failed at the other two.

\begin{figure}[bp]
\centering
\includegraphics[width=\linewidth] {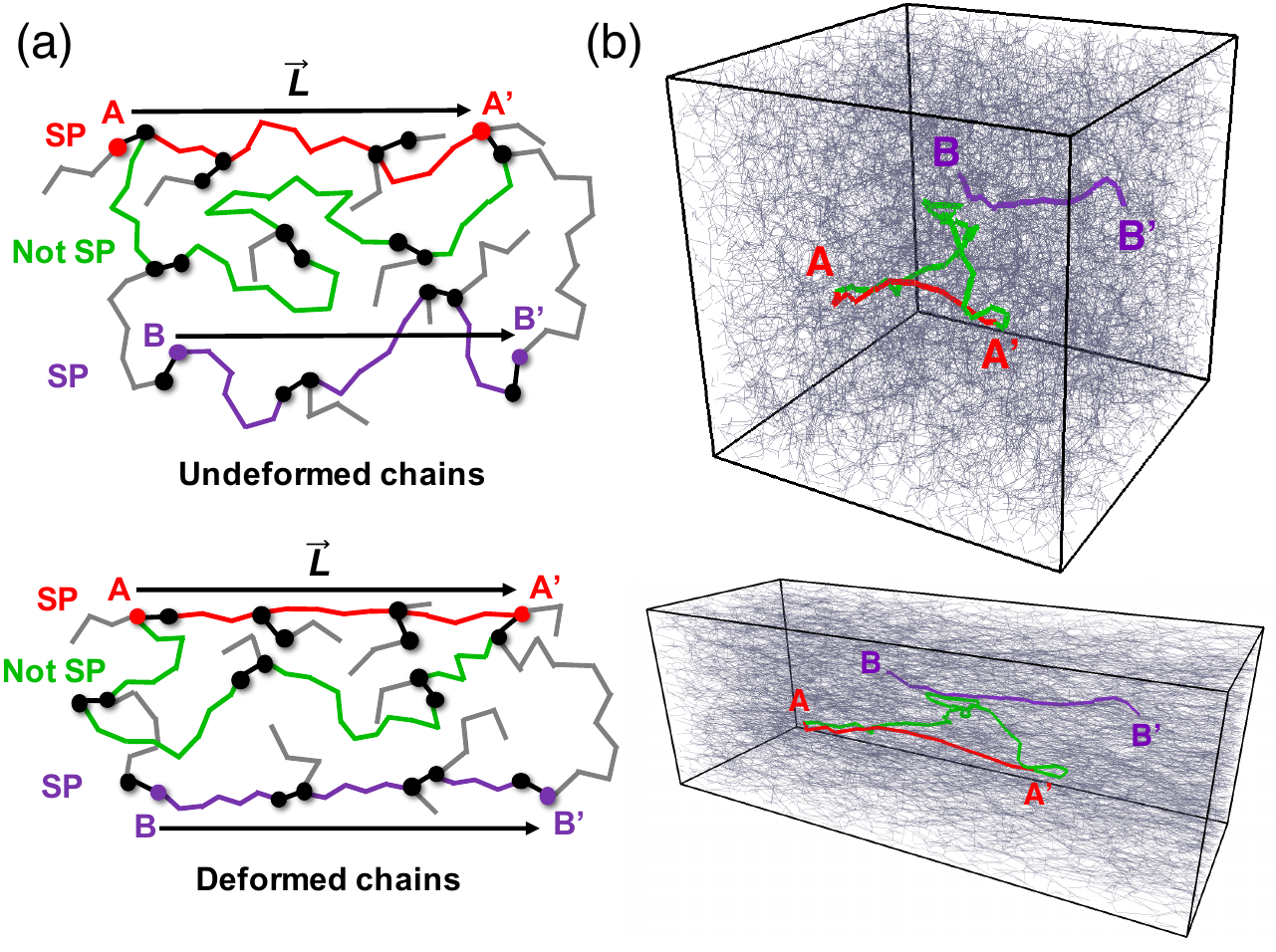}
\caption{(a) Schematic of the polymer network where black line segments indicate cross-links between beads marked with black dots. (b) The undeformed and deformed network of cross-linked beads constructed from CGMD. A-A' and B-B' are example pairs of vertices separated by a vector of $\vec{L}.$ 
Two paths connecting A-A' are colored green and red, and the red path is the SP. The SP connecting B-B' is colored purple.} 
\label{fig:fig3}
\end{figure}

To uncover the controlling microstructural parameter, we perform topological analyses of the elastomer network and examine multiple candidates. We find that local measures such as the lengths of polymer strands~\cite{tehrani2017effect,mark1994elastomeric} between neighboring cross-links on the same molecular chain (referred to as \emph{local chain lengths} below), and the lengths of the shortest macrocycles~\cite{graessley2003polymeric} do not satisfy the conditions given above (see SM). This is because they cannot distinguish between strain-induced and random damage, and are insensitive to the loading directions, i.e. they fail to satisfy criteria (ii) and (iii).

In contrast, by performing SP analyses between \emph{far away} cross-links in the elastomer network as it evolves with deformation, we find that a \emph{global} parameter based on SP lengths does provide the key feature we seek. 
A connection between SPs and limited extensibility of elastomer networks was proposed in~\cite{everaers1996elastic}.
As shown in Fig.~\ref{fig:fig3}, we define a network of cross-linked beads (as vertices) in which every cross-link corresponds to an edge with weight 1.
The network also contains edges between beads connected by backbone chains; the weight of these edges equals the number of backbone bonds between the two beads.
For each bead (e.g. A), we find another bead (e.g. A') that is at a large distance ($\vec{L}$) away from A. 
%
There are multiple paths connecting vertices A and A' on this network, and the length of each path is defined as the sum of the weights of the edges on the path.
The path with the lowest total weight 
is defined as the SP.
While the molecular chains fluctuate in space, the length of the SPs defined above should remain unchanged if there were no bond-breaking events to alter the topology of the network.

\begin{figure*}[tp]
\centering
\includegraphics[width=\linewidth] {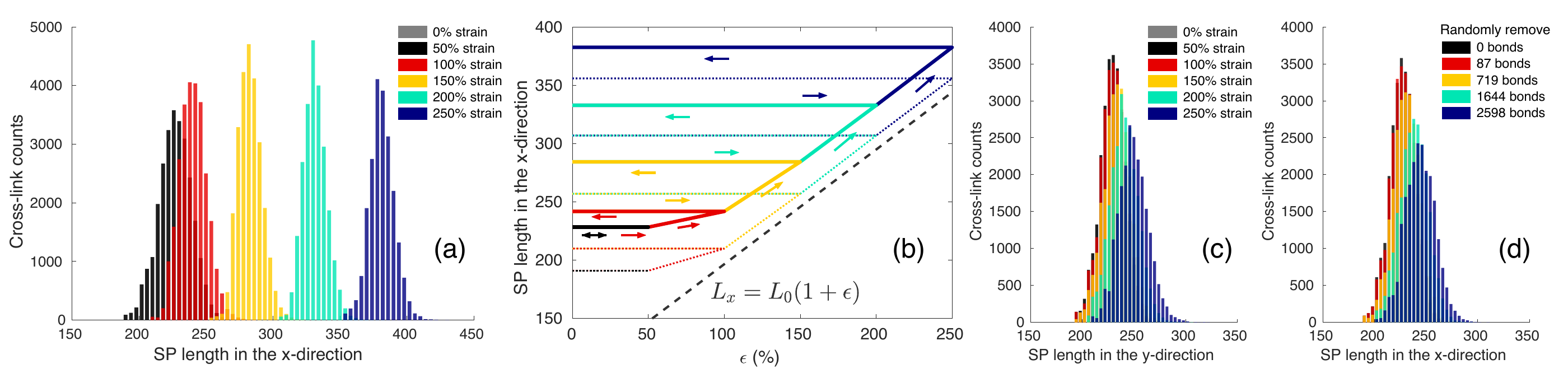}
\caption{(a) Evolution of the SP length distribution in $x$ with straining along $x$. (b) The average SP length (solid line) and minimum SP length (dotted line) in $x$ as functions of $\vec{L_x}$.  
(c) Evolution of the SP length distribution in $y$ with straining along $x$. 
(d) Evolution of the SP length distribution in $x$ for configurations with different amount of cross-links removed randomly. 
} 
\label{fig:fig4}
\end{figure*}


We use Dijkstra's algorithm \cite{dijkstra1959note} to find the SP between each pair of cross-linked beads (on the first network of TN) separated by different distances in the $x$ direction ranging from 0.05 to 1 simulation cell size ($\vec{L_x}$). 
Fig.~\ref{fig:fig4}(a) shows the histograms of all SP lengths between beads separated by $\vec{L_x}$ at different strains.
%
%
%
The histograms at $0\%$ and $50\%$ strain are 
nearly 
the same (Gaussian shape), consistent with the fact that no bond breaking occurs until the strain exceeds $50\%$ (Fig.~\ref{fig:fig1}(c)).

At higher strains, the SP distribution noticeably shifts to longer lengths. 
Remarkably, Fig.~\ref{fig:fig4}(b) shows that the average SP length in the loading ($x$) direction increases almost linearly with strain during loading, and stays constant during unloading, i.e. it exhibits a hysteresis that is fully consistent with the number of broken bonds shown in Fig.~\ref{fig:fig1}(c).  In other words, the SP length distribution satisfies criterion (i) for the microstructural parameter for strain-induced damage.
This behavior can be explained by the requirement that the length of every SP in the $x$ direction (times the physical length of each bond) cannot be shorter than $\vec{L_x}$.  As $\vec{L_x}$ increases linearly with strain (dashed line in Fig.~\ref{fig:fig4}(b)), every SP with length below $\vec{L_x}$ must be decimated by bond breaking.
We confirm that every bond-breaking event occurs on one or more SPs. Destroying these SPs and replacing them by new SPs with longer lengths lead to the shift of histograms. Interestingly, breaking events generally occur on bonds that are traversed by a high number 
of SPs, i.e. on bonds with a high \emph{betweenness centrality} value~\cite{newman2018networks} (see SM). We note that the ratio between the length and the end-to-end separation of a SP is the \emph{tortuosity}, which decreases with strain (see SM). 

To test whether the SP length distribution satisfies criterion (ii), we compute the SP length between each pair of beads separated by the box repeat vector $\vec{L_y}$ in the $y$ direction for unloaded configurations undergone various levels of strain in the $x$ direction.
Fig.~\ref{fig:fig4}(c) shows that straining in $x$ causes only a slight shift of the SP length distribution in $y$, which is very different from the SP length distribution in $x$ shown in Fig.~\ref{fig:fig4}(a).
This confirms that the SP length distribution after strain-induced damage is anisotropic, i.e. criterion (ii) is satisfied.

To test against criterion (iii), we obtain testing configurations by randomly removing cross-links from the undeformed configuration, so that the numbers of broken bonds match those from straining to $100\%$, $150\%$, $200\%$, $250\%$, respectively.
Fig.~\ref{fig:fig4}(d) shows that the resulting SP length distribution (in $x$) appears nearly the same as those shown in Fig.~\ref{fig:fig4}(c), corresponding to SP lengths in $y$ following a loading cycle in $x$.
In other words, as far as SPs in the $y$ direction is concerned, bond breaking caused by straining in the $x$ direction appears indistinguishable from random bond-breaking events.
If the SP length distribution satisfies criterion (iii), we would expect two configurations with the same SP distribution to have the same mechanical response. This is indeed the case shown in Fig.~\ref{fig:fig2}(b), as the stress-strain curve for an elastomer re-loaded in the $y$ direction following a previous loading cycle in $x$ is the same as that for an elastomer with breaking the same number of cross-links randomly.
%

The results above unambiguously demonstrate that the SP length distribution satisfies all the criteria for a suitable microstructural parameter for strain-induced damage in unfilled elastomer with sacrificial bonds.  Given that the distribution is reasonably peaked, the average SP lengths in each direction ($\bar{d}_x$, $\bar{d}_y$, $\bar{d}_z$) may be used as coarse-grained variables characterizing the microstructure of strain-induced damage.
%
%
To serve as the controlling microstructural parameter, the SPs need to be computed between beads that are sufficiently far-away, i.e. to capture the global (instead of local) features of the network topology. For example (see SM), if the separation is less than 0.25$\vec{L_x}$ (about three times the average local chain length at zero strain), then the histograms of SP lengths look significantly different from Fig.~\ref{fig:fig4}(a). 
While the results shown above pertain to TN elastomers, we find the same correspondence between SP lengths and stress-strain hysteresis in SN and DN elastomers, at higher strain where the hysteresis appears. The same behavior is also observed if the backbone bonds, as well as cross-links, can break (see SM).

The majority of previous studies on the mechanical properties of elastomers focused on filled elastomers~\cite{diani2009review,ma2017molecular}, in which the Mullins effect was explained by the breaking of the shortest chains linking two filler particles~\cite{bueche1960molecular}.
Here we show that even without fillers, Mullins effect can arise due to the change of the \emph{global} topology of the elastomer network, as characterized by the SP distributions between far-away cross-links.
Furthermore, we show that the strain-induced damage (in the first network of the multi-network structure) is anisotropic.
The anisotropy of the damage has been observed in filled elastomers~\cite{clough2016covalent,diani2006observation}, but has not been shown experimentally in unfilled elastomers. Our work thus paves the way to the development of new models that accounts for the directional damage in the new unfilled elastomers.

The SP length defined here is very different from the local chain lengths introduced in the \textit{network alteration theory} (NAT)~\cite{marckmann2002theory,chagnon2006development}, i.e.~the length of chain segments between adjacent cross-links on the same chain.
This is a \textit{local} measure of the network structure, while the SP lengths are meant to be applied to well-separated cross-links to measure the \textit{global} property of the network.
Within NAT, bond breaking during deformation causes the average local chain lengths ($N$) to increase and the number of local chains ($n$) to decrease, and such changes are assumed to be responsible for the change of constitutive behavior of the elastomer.
Because $N$ and $n$ defined in NAT are not orientation dependent, they cannot capture the anisotropic nature of strain-induced damage.
In an attempt to capture the anisotropy of damage within the framework of NAT, a generalized definition of local chain lengths has been proposed by replacing $N$ and $n$ with $N_i$ and $n_i$ corresponding to different directions $i$~\cite{diani2006damage}.  
However, $N_i$ and $n_i$ are still not sensitive to the differences between strain-induced and randomly introduced damage (see SM). This is expected for local measures of network topology, given the similar observations on SP lengths if they were computed between cross-links too close to each other.
In addition, our data show that cross-links connected to shorter local chains are not more likely to break (see SM), contrary to what is commonly assumed~\cite{tehrani2017effect,mark1994elastomeric}.

%
The SPs defined in this work is also different from the primitive paths (PPs)~\cite{doi1988theory} introduced to understand the role of entanglements in polymer melts~\cite{everaers2004rheology,li2013challenges}. While a SP follows the (often zigzaged) path of physical bonds, a PP is a (smooth) contour (or tube) that does not necessarily pass through any individual beads or bonds.


In summary, we have used CGMD simulations to understand the strain-induced bond breaking in unfilled multi-network elastomers. The average SP length between far-away cross-links has been identified as the controlling microstructural parameter for damage evolution because it is hysteretic with strain, anisotropic, and it controls the mechanical response. 
Our findings establish a direct connection between the molecular structure and the macroscopic mechanical response of elastomer with sacrificial bonds, and can be used both to develop physics-based models with predictive abilities and to guide the design of new elastomers with improved and targeted mechanical properties.




\bibliography{Submission_PRL.bib}

\end{document}